\documentclass{article}
\usepackage{epsfig,amssymb,amsmath}
\usepackage{dcolumn}% Align table columns on decimal point
\usepackage{bm}% bold math

\begin{document}

\begin{center}
{\LARGE \bf Inductive Reasoning Games as Influenza Vaccination Models: Mean Field Analysis}\vskip.25in
{{\large Romulus Breban, Raffaele Vardavas and Sally Blower}\vskip.1in
{\it Semel Institute for Neuroscience and Human Behavior, University of California, Los Angeles, California 90024, USA.}}
\end{center}
\vskip.35in

\begin{abstract}
We define and analyze an inductive reasoning game of voluntary yearly vaccination in order to establish whether or not a population of individuals acting in their own self-interest would be able to prevent influenza epidemics.  We find that epidemics are rarely prevented.  We also find that severe epidemics may occur without the introduction of pandemic strains.  We further address the situation where market incentives are introduced to help ameliorating epidemics.  Surprisingly, we find that vaccinating families exacerbates epidemics.  However, a public health program requesting prepayment of vaccinations may significantly ameliorate influenza epidemics. 
\end{abstract}\vskip.35in

Game theory has been very successful to help design economic market strategies.  Recently, game theory has been applied in the field of theoretical epidemiology.  Deductive reasoning games have been used to price vaccines \cite{Geo} and to predict the voluntary vaccination coverage (i.e., the proportion of the population that gets vaccinated) for pathogens that provide permanent immunity (smallpox and measles) \cite{Earn1,Earn2}.  However, applying deductive reasoning may be limited \cite{Arthur} because it requires that individuals share the same perception of risk of infection and vaccination adverse effects. For pathogens that provide permanent immunity, using deductive reasoning games is still justified since individuals need to get vaccinated only once.  However, in the case of pathogens that do not provide permanent immunity (e.g., influenza), individuals need to make vaccination decisions every year. It may be assumed that individuals make vaccination decisions based on their past experiences (i.e., use inductive reasoning) rather than based only on the current influenza epidemiology (i.e., use deductive reasoning). Here we present and analyze for the first time several inductive reasoning games that may apply to influenza vaccination.

The first inductive reasoning model was introduced in 1994 with the {\it El Farol bar problem} \cite{Arthur}.  We now briefly present this problem as a paradigm for inductive reasoning games.  Consider a group of $N$ individuals acting in their own self-interest.  Each week, every individual independently decides whether or not to go to the El Farol bar.   An individual deems it worth going if fewer than a critical fraction of the total number of individuals $N$ are present.  Otherwise, the bar is crowded and the individual would rather stay at home. No individual knows the bar attendance in advance, nor do they communicate with others; thus, the decision of going to the bar cannot be made in a logical deductive fashion. Instead, an individual would predict/guess the bar attendance and base their decisions on these predictions. Once the bar attendance is known, an individual evaluates and adapts their predictions for future decisions.  This continuous adaptation of one's behavior based on expectations about future collective behavior is called {\it inductive reasoning} and applies in many instances where logical deductive reasoning fails either in principle or due to bounded rationality \cite{Arthur}.  Inductive reasoning games have been applied to financial markets where traders decide to buy or sell a certain asset whose price is determined by their collective action \cite{ZhangBook,Moro,Challet1,Challet2}. For example, in a market with more buyers than sellers it is rewarding to be in the minority of sellers to benefit from a higher price of the asset.  Such games have been called {\it Minority Games} \cite{ZhangBook,Moro,Challet1,Challet2}. 

We note that there are striking similarities between the El Farol bar problem and the dynamics of the market of voluntary vaccination against pathogens that do not induce permanent immunity.  For concreteness of discussion, we consider influenza vaccination.  The influenza vaccine is effective for one year only, thus individuals must decide every year whether to vaccinate or not \cite{Szucs}.  It may be assumed that individuals act in their own self-interest trying to avoid infection preferably without having to vaccinate. The yearly vaccination coverage (i.e., the proportion of the population vaccinated each year) is determined by the collective participation of the individuals.  Compartmental models of disease transmission \cite{Ferguson1,Cooper,Halloran} have shown that there exists a {\it critical coverage level} such that: if the coverage is below the critical level, an epidemic will occur, otherwise epidemics will be prevented.  However, whether or not a vaccination coverage larger than the critical level is likely to be reached every year in a voluntary vaccination market is a question that we address here for the first time.  Indeed, inductive reasoning games provide a very suitable modeling background for approaching this question.
We construct a inductive reasoning game of influenza vaccination based on the analogy between the situation at hand and the El Farol bar problem.  The critical vaccination coverage will play the role of the critical attendance fraction (i.e., {\it crowding threshold}).  Individuals that go to the bar correspond to individuals that vaccinate.  Individuals that stay at home correspond to individuals that do not vaccinate. For the bar problem: individuals would go to the bar for entertainment, but if the bar is crowded, they would rather stay at home.  For the vaccination game: individuals would vaccinate to avoid infection, but if the coverage is larger than the critical level, they no longer need to vaccinate to avoid infection.  However, there are also distinctions between the two games.  When an individual does not vaccinate and the critical coverage is not reached, they still have a non-zero probability of avoiding infection due to peer vaccination (i.e., they are protected by {\it herd immunity}).  In contrast, for the bar problem, an individual always regrets not going to the bar if the critical attendance is not reached.  We construct an individual-level model that describes the adaptive dynamics of vaccination decisions in a population of non-communicating individuals acting in their own self-interest (i.e., {\it selfish}). Our model tracks both individual-level vaccination decisions and behavior as well as the resulting population-level variables such as influenza prevalence and vaccination coverage levels.  

From the public health point of view, our work has important consequences \cite{Vardavas}.
In the United States, influenza vaccination is voluntary and the demand for influenza vaccines (70 to 75 million vaccine doses per season) is generally met. The vaccine is very effective in offering protection \cite{Beyer}. However, due to evolution of the virus and waning immunity, the vaccine is good only for one influenza season.  In recent years, the vaccination coverage has steadily increased \cite{CDC3,CDC}. Even so, every year, up to 25\% of the United States population is infected with influenza \cite{Szucs} causing 36,000 deaths \cite{CDC,Thompson}. One of the national health objectives of the United States is to further increase the vaccination coverage \cite{CDC3,CDC} which  currently is below the Healthy People 2010 objective \cite{HP2010}.  Therefore, it is important to understand the vaccination dynamics and what incentives could be used to increase the vaccination coverage in order to control the epidemic. 

The outline of the paper is as follows.  In sections \ref{sec:1} we introduce and briefly analyze our individual-level vaccination model with no public health program ({\it basic model}). Our strategy is to derive a dynamical system for the expectations of the variables of the model.  Then, we address deviations from this {\it mean-field} limit by adding small gaussian noise to the dynamical systems. Using the same approach, in section \ref{sec:2} we discuss analytical results regarding the basic model with superimposed vaccination incentives which results into two public health programs.  Our analysis helps explain the coverage dynamics observed in direct numerical simulations of the model.  

\section{Basic Model}
\label{sec:1}

\subsection{Description}

Deterministic models of disease transmission based on ordinary differential equations \cite{Ferguson1,Cooper,Halloran} have shown that there exists a {\it critical coverage level} such that: if the coverage is below the critical level, an epidemic will occur, otherwise epidemics will be prevented.  Our inductive games include a simple model of this coverage threshold\footnote{We do not include the option of treatment. However, the effects of treatment can be implicitly included in our model by decreasing the effective critical vaccination coverage $p_c$.}.  We denote the coverage by $P$, the critical coverage by $p_c$, and the probability of getting infected by $q$.  We consider that if $P<p_c$, then $q(P)$ decreases linearly with $P$, otherwise $q(P)=0$. This model is consistent with the fact that unvaccinated individuals benefit from herd immunity.  Also, it is in qualitative agreement with more sophisticated results from the analysis of the Susceptible-Infected-Recoved (SIR) model; see Appendix \ref{Appendix:A}\footnote{Our model is designed to describe a large population of selfish non-communicative individuals. Thus, we account only for the occurrence of epidemics and we do not consider outbreaks. Outbreaks become decreasingly
 important as the population size $N$ increases.}. The qualitative nature of the results that we describe in this article does not depend on the details of $q(P)$. They only depend on the fact that $q(P)$ is strictly monotonically decreasing for $P<p_c$ and 0 for $P \geq p_c$. We now present the assumptions that define our first inductive reasoning game which we refer to as the {\it basic model}.
\begin{enumerate}
\item We consider a number $N$ of individuals that every year make vaccination decisions.  They are assumed to act in their own interest and not to  communicate their decisions to each other. The sole interest of the individuals is to avoid getting infected, preferably without  getting vaccinated.
\item To make their vaccination decisions, each individual uses their past experiences of vaccination outcomes. Thus, individuals independently decide whether or not to vaccinate using inductive reasoning. 
\item An individual remembers and weights their previous vaccination outcomes with respect to their present vaccination outcome. A parameter $s$ discounts the previous year vaccination outcome with respect to the outcome of the present year ($0\leq s< 1$). For $s=0$, individuals completely ignore the outcome of previous seasons and, as a consequence, do not use inductive reasoning. If $s$ were equal to 1, individuals would not discount the previous vaccination seasons; therefore, the vaccination outcome of the present season is as important as any of the previous seasons.
\item We define a vaccination decision as a Bernoulli variable $\chi^{(i)}_n$ with parameter $w^{(i)}_n$ that depends on a variable $V^{(i)}_n$. $i$ and $n$ are positive integers; $i=1,2,...,N$ labels the individual and $n>0$ labels the seasons. If individual $i$ decided to get vaccinated in season $n$ then $\chi^{(i)}_n=1$, otherwise $\chi^{(i)}_n=0$. $w^{(i)}_n$ is the probability that individual $i$ vaccinates in season $n$.  The variable $V^{(i)}_n$ characterizes the pro-vaccination experience of the $i$th individual (see details in assumption (7)) and determines $w^{(i)}_n$.  The domains of the variables are as follow: $\chi^{(i)}_n\in\{0,1\}$, $w^{(i)}_n\in[0,1]$, and $V^{(i)}_n\in[0,1/(1-s))$.
\item In year $n$, a set $X_n$ of $N$ vaccination decisions is made: $X_n=\{\chi^{(i)}_n; 1\leq i\leq N\}$. Our inductive reasoning game is an array $(X_n)$ where $\{\chi^{(i)}_n, 1\leq i\leq N\}$ determine $\{V^{(i)}_{n+1}; 1\leq i\leq N\}$ which further determine $\{w^{(i)}_{n+1}; 1\leq i\leq N\}$, the parameters of the Bernoulli variables $\{\chi^{(i)}_{n+1}, 1\leq i\leq N\}$ in year $(n+1)$.
\item The infection event of individual $i$ in year $n$ is a Bernoulli variable $\eta^{(i)}_n$ with parameter $q(P_n)$, where $P_n=\sum_{i=1}^N\chi^{(i)}_n/N$ is the coverage achieved that year.  If individual $i$ got infected in season $n$ then $\eta^{(i)}_n=1$, otherwise $\eta^{(i)}_n=0$.
\item $X_n=\{\chi^{(i)}_n; 1\leq i\leq N\}$ and $\{\eta^{(i)}_n; 1\leq i\leq N\}$ determine $V^{(i)}_{n+1}$ as follows (see Fig.~\ref{fig:1}). We have four cases: (a1) if $\chi^{(i)}_n=1$ and $p_c\leq P_n$, then $V^{(i)}_{n+1}=sV^{(i)}_n$; that is, if individual $i$ gets vaccinated in season $n$ and no epidemic occurs, then the individual considers the vaccination unnecessary; (a2) if $\chi^{(i)}_n=1$ and $P_n<p_c$, then $V^{(i)}_{n+1}=sV^{(i)}_n+1$; which means that if individual $i$ vaccinates in season $n$ and an epidemic occurs, then the individual considers the vaccination necessary; (b1) if $\chi^{(i)}_n=0$ and $\eta^{(i)}_n=1$ then $V^{(i)}_{n+1}=sV^{(i)}_n+1$; that is, if individual $i$ does not get vaccinated in season $n$ and gets infected, then the individual considers the vaccination necessary; and (b2) if $\chi^{(i)}_n=0$ and $\eta^{(i)}_n=0$ then $V^{(i)}_{n+1}=sV^{(i)}_n$; which means that if individual $i$ get vaccinated in season $n$ and they do not get infected, then the individual considers the vaccination unnecessary.
\item The probability that an individual chooses to get vaccinated is updated as follows \begin{eqnarray} w^{(i)}_{n+1}=V^{(i)}_{n+1}/[(s^{n+1}-1)/(s-1)].\label{eq:w}\end{eqnarray} 
That is, an individual's probability to get vaccinated in the next season is given by the updated cumulative vaccination experience. We have normalized $V^{(i)}_{n+1}$ by $(s^{n+1}-1)/(s-1)$ because this factor is the maximum possible value for $V^{(i)}_{n+1}$ if individual $i$ would have benefited from vaccination in all of the $n$ influenza seasons.
\end{enumerate}
We consider initial conditions that assign  a random vaccination probability for the first season to every individual. Specifically, $V^{(i)}_0=0$ and $w^{(i)}$ is uniformly random between $0$ and $1$. Our initial conditions are chosen to reflect the fact that the initial public awareness of the benefits of the influenza vaccination is not high enough to prevent an epidemic, while at the individual-level the likelihood of vaccination may vary considerably. Figure \ref{fig:2} shows numerics obtained by simulating the basic model. Our analysis aims to provide an understanding of the coverage dynamics in Fig.~\ref{fig:2}. As $p$ approaches $p_c$ from below, it eventually fluctuates above $p_c$, then abruptly drops below $p_c$, and the dynamics repeats.

\subsection{Analysis}

We first present an analysis of our featured model in the limit of large population. Then, we discuss asymptotic aspects of the model that apply as the population is large, yet finite. We proceed with the derivation of a one-dimensional iterated map in the variable $p$ denoting the average of $w^{(i)}$ over the entire population.  Note that $w^{(i)}_n=E(\chi^{(i)}_n)$, $p_n=E(P_n)=\sum_{i=1}^Nw^{(i)}_n/N$, and $E(q(P))=q(p)$ where by $E()$ we denote expectation. Following the tree in Fig.~\ref{fig:1} which describes the four cases of vaccination evaluation, we obtain
\begin{eqnarray}
\begin{array}{l|l|l}
\mbox{branch}&\mbox{expected population fraction}&\mbox{$V^{(i)}$ update}\\\hline
\mbox{(a1,2)}&p_n&V^{(i)}_{n+1}=s V^{(i)}_n+1-\theta(p_n-p_c);\\
\mbox{(b1)}&(1-p_n)q(p_n)&V^{(i)}_{n+1}=s V^{(i)}_n+1;\\
\mbox{(b2)}&(1-p_n)[1-q(p_n)]&V^{(i)}_{n+1}=s V^{(i)}_n;\\
\end{array}
\label{eq:V}
\end{eqnarray} 
where $\theta(x)$ is the unit step function defined as
\begin{eqnarray}
\theta(x)=\left\{\begin{array}{ll}1, & \mbox{if $x\geq 0$};\\0, & \mbox{if $x< 0$};\end{array}\right.
\end{eqnarray}
and $q(p)$, the probability of an unvaccinated individual getting infected with influenza is given by
\begin{eqnarray}
q(p)=\left\{\begin{array}{ll}0, & \mbox{if $p\geq p_c$};\\-p\, q(0)/p_c+q(0), & \mbox{if $p<p_c$}.\end{array}\right.
\end{eqnarray} 
Taking the weighted average over Eqs.~\eqref{eq:V}, we obtain
\begin{eqnarray}
v_{n+1}=s v_n+(1-p_n)q(p_n)+p_n[1-\theta(p_n-p_c)],
\label{eq:v}
\end{eqnarray}
where $v$ denotes the average of $V^{(i)}$ over the entire population.
Taking the population average over Eqs.~\eqref{eq:w}, we get
\begin{eqnarray}
p_{n+1}=(1-s) v_{n+1}/(1-s^{n+1}).
\label{eq:p}
\end{eqnarray}
Combining Eqs.~\eqref{eq:v} and \eqref{eq:p}, we describe the dynamics of the vaccination coverage at the population level in the limit of infinite population, without regards to the individual-level processes 
\begin{eqnarray}
p_{n+1}=\frac{s(1-s^n)}{1-s^{n+1}} p_n+\frac{1-s}{1-s^{n+1}}\{(1-p_n)q(p_n)+p_n[1-\theta(p_n-p_c)]\}.
\label{eq:p_}
\end{eqnarray}
Our dynamical system is defined on the unit interval $\mathcal{I}=[0,1]$.

\subsubsection{Fixed Point Analysis}
\label{sec:FPA}
The fixed points of our map are determined by the following autonomous asymptotic form obtained from Eq.~\eqref{eq:p_} as $n\rightarrow\infty$
\begin{eqnarray}
p_{n+1}=s p_n+(1-s)\{(1-p_n)q(p_n)+p_n[1-\theta(p_n-p_c)]\}.\label{eq:asymptote}
\end{eqnarray}
Due to the discontinuities at $p=p_c$, we distinguish two complementary domains: $\mathcal{I}_1=[p_c, 1]$ and $\mathcal{I}_2=[0, p_c)$.

{\bf Case 1: $\mathcal{I}_1$.} Equation \eqref{eq:asymptote} becomes the following linear dynamical system
\begin{eqnarray}
p_{n+1}=s p_n.
\label{eq:asymptote1}
\end{eqnarray}
The above dynamical system has no attractors in $\mathcal{I}_1$.  However, if we extend the domain to $\mathcal{I}$, then the system has a fixed point at $p^*=0$ which is a global attractor that belongs to $\mathcal{I}_2$.  This fully characterizes the dynamics in $\mathcal{I}_1$: orbits in $\mathcal{I}_1$ will be attracted to $0$ until they land in $\mathcal{I}_2$. In $\mathcal{I}_2$ the orbit is iterated with a different smooth map.

{\bf Case 2: $\mathcal{I}_2$.} We now obtain the following nonlinear dynamical system
\begin{eqnarray}
p_{n+1}=p_n+(1-s)(1-p_n)(-p_n\,q(0)/p_c+q(0)),
\label{eq:asymptote2}
\end{eqnarray}
that has no fixed points in $\mathcal{I}_2$. However, if we extend the domain to $\mathcal{I}$, then the system has two fixed points: $\tilde p=1$ and $\hat p=p_c$. The derivative of the map evaluated at $\tilde p$ is $1+q(0)(1-s)(p_c^{-1}-1)>1$; thus, this fixed point is unstable for all the parameter values. $\hat p$ is a potential attractor of the system since the derivative of the map evaluated at $\hat p$ is $1-q(0)(1-s)(p_c^{-1}-1)$ with range $(-\infty,1]$. It is important to note that $\hat p$ lies on the boundary between $\mathcal{I}_1$ and $\mathcal{I}_2$. Thus, even though $\hat p$ does not belong to $\mathcal{I}_2$, $\hat p$ attracts orbits with initial conditions in $\mathcal{I}_2$. That is, the basin of attraction of $\hat p$ intersects $\mathcal{I}_2$. Assuming that $\hat p$ is an attractor and derivative of the map at $\hat p$ is nonnegative, orbits that start in $\mathcal{I}_2$ are immediately attracted from below to $\hat p$, but never reach this fixed point. They only get arbitrarily close to it, always remaining in $\mathcal{I}_2$. Orbits that start in $\mathcal{I}_1$ are initially attracted to $p^*=0$ but once they land in $\mathcal{I}_2$ they are attracted to $\hat p$. For the parameters of the numerics presented in Fig.~\ref{fig:2}, $\hat p$ is an attractor and the derivative of the map at $\hat p$ is nonnegative.

\subsubsection{Bifurcation Diagram}
\begin{figure}
\begin{center}
\mbox{\epsfig{file=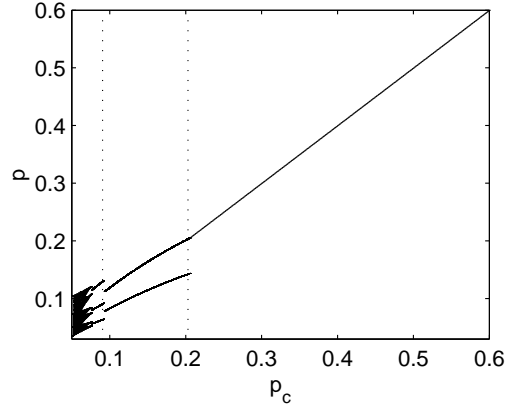,width=6.8cm}}\end{center}
\caption{Bifurcation diagram of the dynamical system given by Eq.~\eqref{eq:p_} versus $p_c$. $s=0.7$ and the probability of getting infected when the coverage $p=0$ is $q(0)=0.8$.  We note that numerical noise greatly perturbs the dynamics of our piecewise smooth map because $\hat p \in \partial\mathcal{I}_2$. To improve the numerics, we slightly modified the map in order for $\hat p \in \mathcal{I}_2$; in turn, this slightly changes the bifurcation values of $p_c$.}
\label{fig:2.6}
\end{figure}

With varying $p_c$ and/or $s$ and/or $q(0)$, a rich dynamical behavior is observed for our iterated map.  As expected in piecewise smooth systems, we observe {\it border-collision bifurcations} \cite{NOY,BBC,DNOYY,LM,NY,BFHH,BBC2}. A bifurcation diagram for $s=0.7$, $q(0)=0.8$ and varying $p_c$ is presented in Fig.~\ref{fig:2.6}.  With decreasing $p_c$, the derivative of the map at $\hat p$, $1-(1-s)[q(0)(p_c^{-1}-1)]$, becomes zero at $p_c=\{1+1/[(1-s)q(0)]\}^{-1}\approx 0.2$.  At this critical $p_c$ value, a border collision bifurcation takes place and a stable periodic orbit is created in phase space.  Approaching $\hat p$ for $p_c<\{1+1/[(1-s)q(0)]\}^{-1}$, an orbit in $\mathcal{I}_2$ will land in $\mathcal{I}_1$ since the derivative of the map at $\hat p$ is negative.  The smooth map that applies to the $\mathcal{I}_1$ domain [Eq.~\eqref{eq:asymptote1}] will send the orbit back in $\mathcal{I}_2$, and, eventually, a stable period two orbit is created.  With further decreasing $p_c$, $\hat p$ becomes unstable at $p_c=\{1+2/[(1-s)q(0)]\}^{-1}\approx 0.1$ and the period two orbit undergoes period doubling.  At low values of $p_c$, chaotic behavior is numerically observed.

\subsubsection{Effects at Large Finite $N$}
\label{sec:noise}
\begin{figure}
\begin{center}
\mbox{\epsfig{file=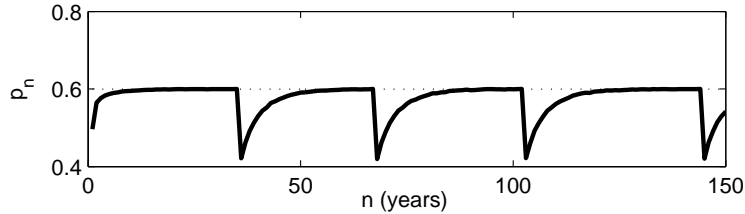,width=10cm}}\end{center}
\caption{Vaccination dynamics using a memory parameter $s=0.7$, a critical coverage $p_c=0.6$ (dashed line) and a probability $q(0)=0.8$ of getting infected when the coverage $p=0$. Dynamics of yearly coverage $p$ for a population of $N= 10^5$ individuals. The dynamics of $p$ is approximately cyclic: as $p$ approaches $p_c$ from below, it eventually fluctuates above $p_c$ and then abruptly drops below $p_c$.} 
\label{fig:2}
\end{figure}

We now discuss several aspects of the dynamics that occur because the number of individuals $N$ is in fact finite, albeit large.  These aspects are transparent in our direct simulations of the game and the above fixed point analysis is insufficient to explain the results in Fig.~\ref{fig:2}. 

From the point of view of the mean field analysis, the case of large finite $N$ can be described by adding small amplitude gaussian noise to the mean field map.  In most of the phase space the noise does not change the qualitative dynamics of the orbit.  However, the situation becomes critically different as the noisy orbit asymptotically approaches $\hat p$ from $\mathcal{I}_2$. Due to the noise (i.e., stochasticity in the mean field due to the finite number of individuals), $p$ may jump above (but close to) $\hat p$. According to Eq.~\eqref{eq:asymptote1}, in the next iteration the orbit lands in $\mathcal{I}_2$ in the vicinity of $s p_c$, far from $\hat p=p_c$; see Fig.~\ref{fig:2}.  Then, the orbit is attracted again to $\hat p$ and undergoes a similar scenario in an apparently periodic dynamics.  This phenomenon may be called {\it noise induced border-collision bifurcation} since the presence of arbitrarily small noise transforms an orbit asymptotically approaching $\hat p$ into an orbit that is expected to be periodic.  We note that this sensitivity of the mean field of the basic model to arbitrary small noise makes the simulation of the mean field of the basic model difficult since rounding noise cannot be avoided in numerics.

The expected periodicity depends on the number of individuals in the ensemble and can be estimated as follows.  Close to $\hat p$, the dynamics of $p$ can be approximated as $\hat p-p_n\sim \lambda^n$, where $\lambda$ is the derivative of the map at $\hat p$; for the parameter values used in Fig.~\ref{fig:2}, $\lambda=1-(1-s)[q(0)(p_c^{-1}-1)]\approx 0.84$.  Denoting $N_n^v$ and $\hat N$ the numbers of individuals that vaccinate in season $n$ and the number of individuals that would vaccinate at equilibrium, the previous relation can be written as $\hat N-N_n^v\sim\lambda^nN$. We expect a jump of the coverage $p$ above $\hat p$ when $\hat N-N_n^v\sim\hat N^{1/2}$ which is the order of the fluctuations in $N_n^v$.  Denoting $\tilde n$ the expected period of the dynamics, we obtain the following scaling result at large finite $N$
\begin{eqnarray}
\tilde n\sim -\frac{\log (N p_c)}{2 \log\lambda},
\label{eq:period}
\end{eqnarray}
which we have successfully verified through numerics (not shown).   

\section{Models with Public Health Programs}
\label{sec:2}

The basic model predicts that epidemics are not prevented.  Furthermore, severe epidemics are periodically expected.  We explore two possible public health strategies that may ameliorate epidemics. Thus, we introduce two additional inductive reasoning games in order to evaluate the potential effects of the following two public health programs applied to the basic model:	
\begin{itemize}
	\item{\bf Program 1: }{If the head of the family (HF) pays to get vaccinated then their family will get vaccinated for free.}
	\item{\bf Program 2: }{If an individual pays to get vaccinated then that individual will get free vaccinations for a specified number of successive years.}
\end{itemize}
We follow our previous strategy, deriving and analyzing mean-field approximations. Then, we discuss first order deviations from the mean-field exploring the effects of superimposed small noise.  

\subsection{Public Health Program 1}
\subsubsection{Model Description}
We consider that the population of $N$ individuals is now divided into $F$ groups representing families. Each family contains $C$ members and one individual in each family acts as its head.  The public health program offers free vaccination to a family if the head of that family paid for his/her vaccination.  Only the heads of the family make vaccination decisions and they track the vaccination experience for all of their family members; see Fig.~\ref{fig:family}. It is very important to note that, as a consequence of this public health program, the vaccination coverage of the heads of families equals the population-level vaccination coverage. We specify the model using a set of eight assumptions. 
\begin{enumerate}
\item We consider a number of $F=N/C$ HF that every year make vaccination decisions.  They are assumed to act in the interest of their own families (including themselves) and not to communicate their decisions to each other. The sole interest of the HF is to protect their family members from infection, preferably without getting anyone vaccinated.
\item To make their vaccination decisions, each HF uses inductive reasoning. 
\item A parameter $s$ discounts the HF's previous year vaccination outcome with respect to the outcome of the present year ($0\leq s< 1$). 
\item A vaccination decision is a Bernoulli variable $\chi^{(i)}_n$ with parameter $w^{(i)}_n$ that depends on a variable $V^{(i)}_n$. $i$ and $n$ are positive integers; $i=1,2,...,F$ labels the HF and $n>0$ labels the seasons.  If HF $i$ decided to vaccinate in season $n$ then their family also gets vaccinated and $\chi^{(i)}_n=1$; otherwise, $\chi^{(i)}_n=0$. $w^{(i)}_n$ is the probability that HF $i$ vaccinates in season $n$. $V^{(i)}_n$ is the pro-vaccination experience of the family of HF $i$ (see assumption (7)) and determines $w^{(i)}_n$.  The domains of the variables are as follow: $\chi^{(i)}_n\in\{0,1\}$, $w^{(i)}_n\in[0,1]$, and $V^{(i)}_n\in[0,C/(1-s))$.
\item In year $n$, a set $X_n$ of $F$ vaccination decisions is made: $X_n=\{\chi^{(i)}_n; 1\leq i\leq F\}$. Our inductive reasoning game is an array $(X_n)$ where $\{\chi^{(i)}_n, 1\leq i\leq F\}$ determine $\{V^{(i)}_{n+1}; 1\leq i\leq F\}$ which further determine $\{w^{(i)}_{n+1}; 1\leq i\leq F\}$, the parameters of the Bernoulli variables $\{\chi^{(i)}_{n+1}, 1\leq i\leq F\}$ in year $(n+1)$.
\item The infection event of individual $j$ ($j=1,...,N$; i.e., the individual may or may not be HF) in year $n$ is a Bernoulli variable $\eta^{(j)}_n$ with parameter $q(P_n)$, where $P_n=\sum_{i=1}^F\chi^{(i)}_n/F$ is the coverage achieved that year.  If individual $j$ got infected in season $n$ then $\eta^{(j)}_n=1$, otherwise, $\eta^{(j)}_n=0$.
\item $X_n=\{\chi^{(i)}_n; 1\leq i\leq F\}$ and $\{\eta^{(i)}_n; 1\leq i\leq F\}$ determine $V^{(i)}_{n+1}$ as follows (see Fig.~\ref{fig:1}). We have $C+3$ cases: (a1) if $\chi^{(i)}_n=1$ and $p_c\leq P_n$, then $V^{(i)}_{n+1}=sV^{(i)}_n$; that is, if HF $i$ gets their family (including themselves) vaccinated in season $n$ and no epidemic occurs, then the HF considers the vaccination unnecessary; (a2) if $\chi^{(i)}_n=1$ and $P_n<p_c$, then $V^{(i)}_{n+1}=sV^{(i)}_n+C$; which means that if HF $i$ gets their family (including themselves)  vaccinated in season $n$ and an epidemic occurs, then the HF considers the vaccination necessary for all the family members; (bk) if $\chi^{(i)}_n=0$ and $k$ family members ($k=0, ..., C$) have $\eta^{(j_i)}_n=1$ (where $j_i=0, ..., C$ labels the family members of the $i$th family), then $V^{(i)}_{n+1}=sV^{(i)}_n+k$; that is, if HF $i$ does not get vaccinated in season $n$ and $k$ members of their family (including themselves) get infected, then the HF adjusts his pro-vaccination experience by accounting for their infected family members.
\item The probability that an HF chooses to vaccinate is updated as follows \begin{eqnarray} w^{(i)}_{n+1}=V^{(i)}_{n+1}/[C(s^{n+1}-1)/(s-1)].\label{eq:ww}\end{eqnarray} 
That is, an HF's probability to get vaccinated in the next season is given by updated cumulative vaccination experience. We have normalized $V^{(i)}_{n+1}$ by $C(s^{n+1}-1)/(s-1)$ because this factor is the maximum possible value for $V^{(i)}_{n+1}$ if HF $i$ and their family would have benefited from vaccination in all of the $n$ influenza seasons.
\end{enumerate}
Figure \ref{fig:programs}A shows numerics obtained by simulating the model with the first public health program.  We note that the coverage drops below $p_c$ more often, and thus the first public health program appears detrimental to ameliorating epidemics. Our analysis aims to provide an understanding of the coverage dynamics in Fig.~\ref{fig:programs}A. 

\subsubsection{Model Analysis}

\begin{figure}
\begin{center}
\mbox{\epsfig{file=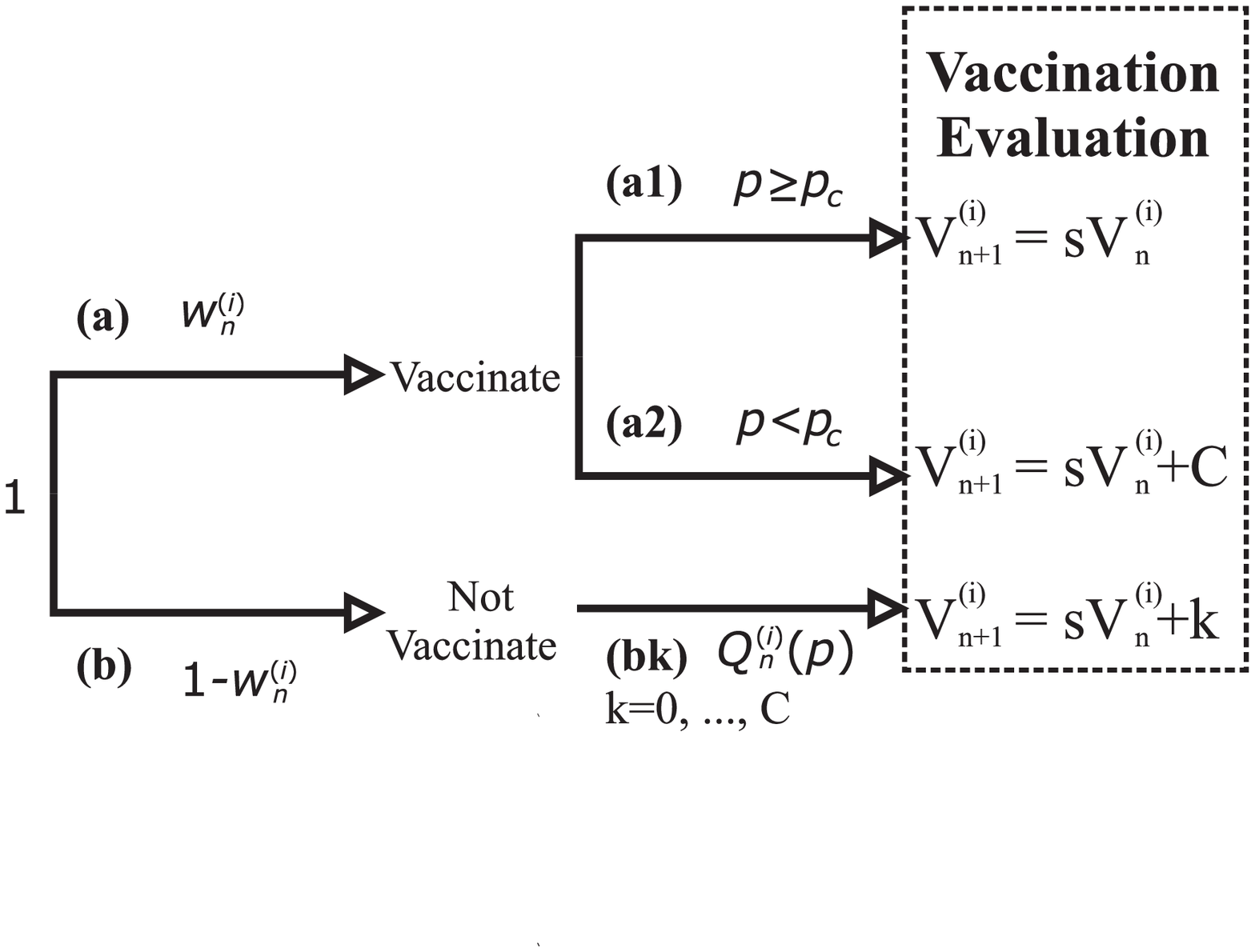,width=13cm}}\end{center}
\caption{Schematics illustrating the evaluation tree for each head of family. An head that decides to vaccinate themselves and their family (branch (a)) will judge their choice depending on whether there was an epidemic that season.  If the coverage was equal or greater than the critical coverage $p\geq p_c$ (branch (a1)), they will conclude that their choice to get vaccinated that season was not necessary to prevent infection. Otherwise, if the coverage was lower than the critical coverage $p<p_c$ (branch (a2)), they will conclude that their choice was beneficial for avoiding infection that season. An head that decides not to vaccinate themselves and their family (branch (b)) will judge their choice based on how many of their family members were infected. If $k$ members get infected (branch (bk)), they will conclude that vaccination was necessary only for $k$ members of the family.}
\label{fig:family}
\end{figure}

Following the HF evaluation tree shown in Fig.~\ref{fig:family} we obtain
\begin{eqnarray}
\hspace{-7mm}\begin{array}{l|l|l}
\mbox{branch}&\mbox{expected population fraction}&\mbox{$V^{(i)}$ update}\\\hline
\mbox{(a1,2)}&p_n&V^{(i)}_{n+1}=s V^{(i)}_n+C[1-\theta(p_n-p_c)];\\
\mbox{(bk)}&(1-p_n)Q_k(p_n)&V^{(i)}_{n+1}=s V^{(i)}_n+k;\\
\end{array}
\label{eq:Vfamily}
\end{eqnarray}
where $Q_k(p)$ is the probability that $k$ members get infected with influenza in an unvaccinated family when the expected coverage is $p$. The probability that a single individual gets infected in a season with expected coverage $p$ is $q(p)$. Assuming mass-action infection dynamics \cite{McCallum}, the probability that $k$ members of a family get infected is binomial
\begin{eqnarray}
	Q_k(p_n) = \frac{C!}{k!(C-k)!}q(p)^k\left [ 1-q(p) \right ]^{C-k}.
	\label{eq:binomial}	
\end{eqnarray}

We now present mean-field equations for our model with the first public health program. Substituting Eq.~\eqref{eq:binomial} into Eq.~\eqref{eq:Vfamily} and averaging over all branches we obtain
\begin{eqnarray}
v_{n+1}=s v_n+Cp_n[1-\theta (p_n-p_c)]+C q(p_n) (1-p_n).
\end{eqnarray}
Averaging over Eq.~\eqref{eq:ww} we obtain
\begin{eqnarray}
p_{n+1}=v_{n+1}/[C (s^{n+1}-1)/(s-1)].
  \label{eq:pfamily}	
\end{eqnarray}

The fixed point analysis of the above dynamical system follows similarly to that of the basic model and yields similar results. As in the case of the basic model, $\hat p=p_c$ plays a crucial role in determining the dynamics.  The derivative of the coverage map at $\hat p$ is the same as in the basic model.  The expected periodicity formula is similar to that found for the basic model [Eq.~\eqref{eq:period}]. In this case the number of HF ($F=N/C$) determines the periodicity instead of the total number of individuals ($N$). Therefore, the periodicity as a function of the family size $C$ can be expressed as
\begin{eqnarray}
\tilde n(C)\sim -\frac{\log (F p_c)}{2 \log\lambda},
\end{eqnarray}
where for $C=1$ we recover Eq.~\eqref{eq:period}. We now compare the expected periodicity of major epidemics for $C>1$ to that for $C=1$ (i.e., basic model). We use the same values of $s$, $p_c$ and $q(0)$ such that orbits in both models approach $\hat p$ from $\mathcal{I}_1$.  The ratio of the expected periodicities is given by
\begin{eqnarray}
\tilde n(C)/\tilde n(1)\sim  \left [1-\frac{\log C}{\log (N p_c)}\right]<1.
\end{eqnarray}
Hence, the first public health program increases the frequency of major influenza epidemics and decreases the time average of the coverage. 

\subsection{Public Health Program 2}
\subsubsection{Model Description}
\begin{figure}
\begin{center}
\mbox{\epsfig{file=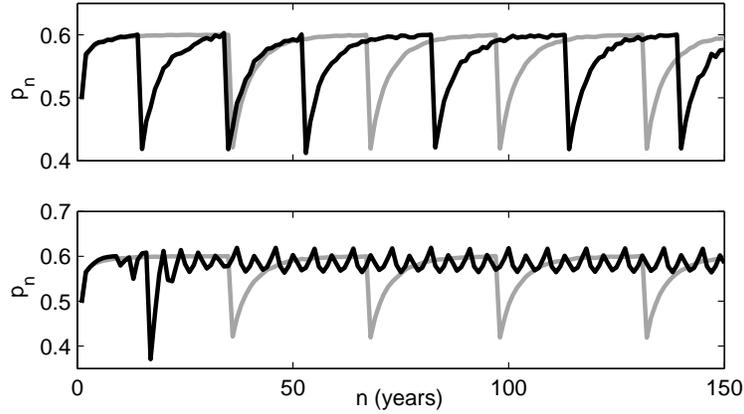,width=10cm}}\end{center}
\caption{Coverage dynamics ($p$) for different public heath programs in a population of $N=10^5$ non-communicating selfish individuals using a memory parameter $s=0.7$, a critical coverage $p_c=0.6$, and a probability $q(0)=0.8$ of getting infected when the coverage $p=0$. \textbf{A}: The head of the family makes the decision as to whether or not their family vaccinates. The coverage dynamics when the family size is eight ($C=8$) is shown in black; the coverage dynamics when individuals make vaccination decisions independently is shown in gray for comparison. Similar results were obtained for family sizes of two and four. \textbf{B}: Individuals that pay for one vaccination are rewarded $y=3$ extra years of vaccination; the coverage dynamics is shown in black and the time averages of the epidemiological quantifiers are ameliorated. The coverage dynamics when individuals pay for every year of vaccination is shown in gray for comparison.} 
\label{fig:programs}
\end{figure}

An individual who pays to enter the second public health program receives influenza vaccination for the current year and for $y$ number of successive years. Although individuals enrolled in the program do not make vaccination decisions, they consider the necessity of vaccination every year.  At the end of   the free vaccination period, they use their evaluations to decide whether to pay for another enrollment.  The model can be formally described by a number of eight assumptions.  In particular, we use assumptions (1) through (7) of the basic model.  Assumption (8) is the following.
\begin{enumerate}
\item[(8)] The probability that an individual gets vaccinated is updated as \begin{eqnarray} w^{(i)}_{n+1}=V^{(i)}_{n+1}/[(s^{n+1}-1)/(s-1)],\label{eq:w2}\end{eqnarray}
unless $\chi^{(i)}_n=1$; i.e., the individual vaccinated in season $n$.  In this later case, $w^{(i)}_{n+r}=1$ for $0<r\leq y$, and, in season $(n+y+1)$, \begin{eqnarray} w^{(i)}_{n+y+1}=V^{(i)}_{n+y+1}/[(s^{n+y+1}-1)/(s-1)].\label{eq:w3}\end{eqnarray}
That is, after vaccinating in season n, and taking advantage of $y$ seasons of free vaccination, the individual resumes his adaptive behavior in season $(n+y+1)$.
\end{enumerate}
Figure \ref{fig:programs}B shows numerics obtained by simulating the model with the second public health program.  We note a qualitative change in the orbit of $p$; thus, the second public health program can potentially ameliorate epidemics. Our analysis aims to provide an understanding of the coverage dynamics in Fig.~\ref{fig:programs}B. 
\subsubsection{Model Analysis}
We now introduce notation to describe the analysis of the model. We use the superscript $r$ to specify the individual-level parameters for individuals that have $r$ vaccinations left. $N_n^r$ ($0\leq r \leq y$)  denotes the number of individuals that have $r$ vaccination years left in season $n$; $\sum_{r=0}^{y} N_n^r=N$.  The ratio between the number of individuals with $r$  vaccination years that vaccinate in season $n$ and $N_n^r$ is denoted by $\pi_n^r$. Since we assume that all individuals complete the free vaccination program, $\pi_n^r=1$ if $r>0$. The population-level coverage can be written as
\begin{eqnarray}
	p_n= \frac{1}{N} \sum_{r=0}^y N^r_n \pi_n^r.
	\label{eq:pyear}
\end{eqnarray}
Given $\pi_n^0$ we can write a dynamical system for $N_n^r$
\begin{eqnarray}
	N_{n+1}^0&=&N_n^1+(1-\pi^0_n)N^0_n,\nonumber\\
	N_{n+1}^r&=& N_{n}^{r+1}; \,\, \mbox{for $0 <r<y$}, \\
	N_{n+1}^y&=& \pi^0_n N^0_n.\nonumber
	\label{eq:Nyear}
\end{eqnarray}
The number of individuals that will decide to enroll in the program next year ($N_{n+1}^0$) is given by the individuals that will finish their vaccination program ($N_n^1$) and the individuals that did not enroll in the vaccination program this year ($(1-\pi^0_n)N^0_n$). The number of individuals with $r$ ($0<r<y$) years left in the program next year  ($N_{n+1}^r$) is given by the number of individuals with $r+1$ years left in the program this year. The number of individuals with $y$ years left in the program  next year ($N_{n+1}^y$) is given by the number of individuals that enroll in the vaccination program this year ($\pi^0_n N^0_n$).  

\begin{figure}
\begin{center}
\mbox{\epsfig{file=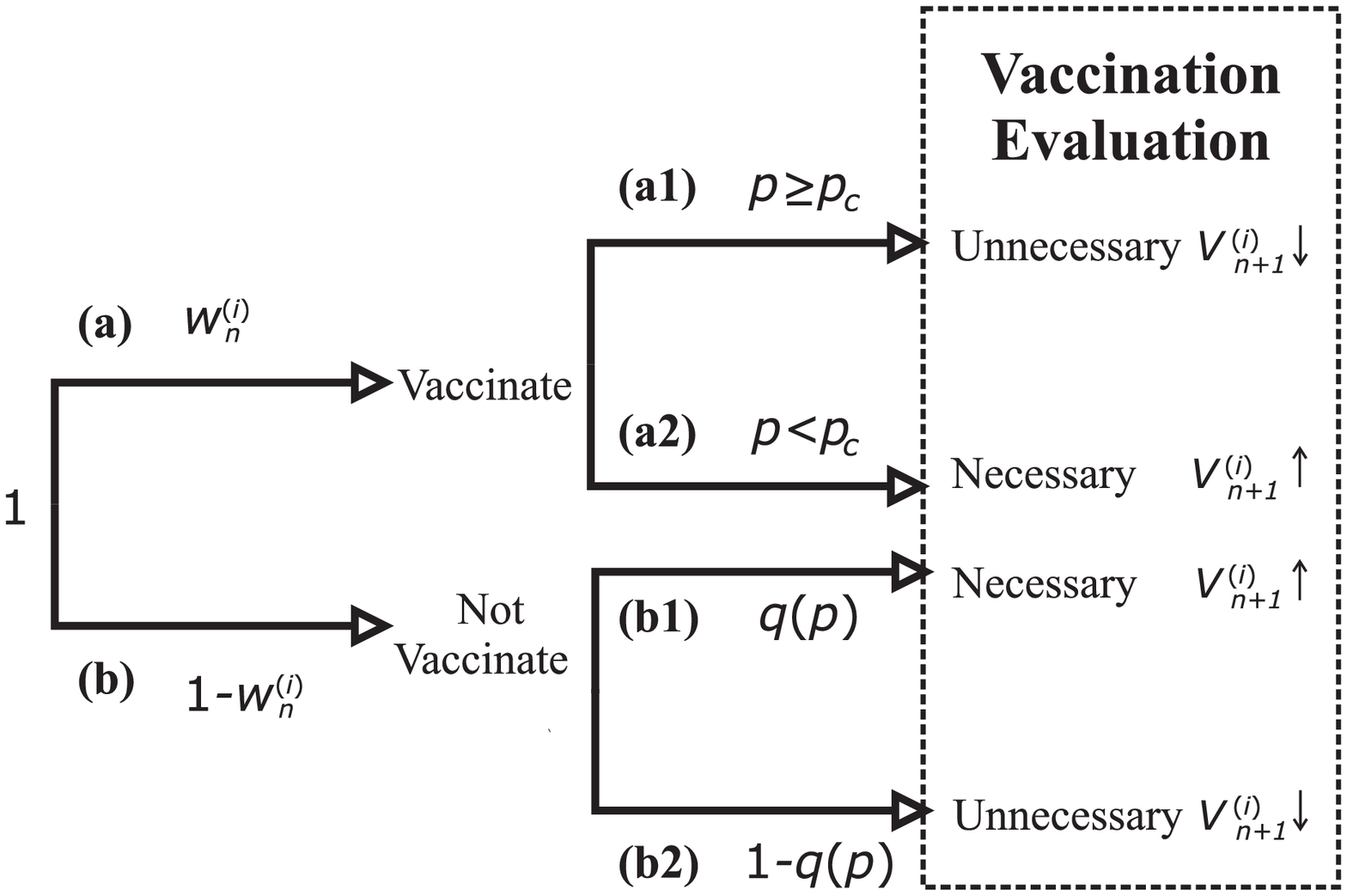,width=13cm}}\end{center}
\caption{\textbf{A}: A schematic diagram illustrating the evaluation tree for each selfish individual. An individual that decides to vaccinate (branch (a)) will judge their choice depending on whether there was an epidemic that season.  If the coverage was equal or greater than the critical coverage $p\geq p_c$ (branch (a1)), they will conclude that their choice to get vaccinated that season was not necessary to prevent infection. Otherwise, if the coverage was lower than the critical coverage $p<p_c$ (branch (a2)), they will conclude that their choice was beneficial for avoiding infection that season. An individual that decides not to vaccinate that season (branch (b)) will judge their choice based on whether they was infected. If they did get infected (branch (b1)) they will conclude that their choice of not vaccinating was detrimental and that vaccination was necessary for avoiding infection. Instead, if by chance they avoided infection (branch (b2)), they will conclude that vaccination was not necessary.}
\label{fig:1}
\end{figure}

The vaccination behavior of the individuals enrolled in the vaccination program is simple: they vaccinate every year 
\begin{eqnarray}
	w^{r\,(i)}_n=1.
\end{eqnarray}
However, they evaluate the necessity of vaccination updating their pro-vaccination variable $V$ depending on whether or not there was an influenza epidemic each year
 \begin{eqnarray}
	V^{r\,(i)}_{n+1}=s V^{r\,(i)}_{n}+1-\theta(p_n-p_c).
\end{eqnarray}    
The individuals that are not enrolled in the vaccination program need to decide whether or not to vaccinate. The individual-level probabilities to vaccinate get updated  as in the basic model
\begin{eqnarray}
w^{0\,(i)}_{n+1}=V^{0\,(i)}_{n+1}/[(s^{n+1}-1)/(s-1)].
\end{eqnarray} 
The evaluation of the vaccination decisions is the same as in the basic model; see Fig.~\ref{fig:1}. Following the evaluation tree in Fig.~\ref{fig:1} we obtain
\begin{eqnarray}
\begin{array}{l|l|l}
\mbox{branch}&\mbox{population fraction}&\mbox{$V^{0\,(i)}$ update}\\\hline
\mbox{(a1,2)}&\pi^0_n&V^{0\,(i)}_{n+1}=s V^{0\,(i)}_n+1-\theta(p_n-p_c);\\
\mbox{(b1)}&(1-\pi^0_n)q(p_n)&V^{0\,(i)}_{n+1}=s V^{0\,(i)}_n+1;\\
\mbox{(b2)}&(1-\pi^0_n)(1-q(p_n))&V^{0\,(i)}_{n+1}=s V^{0\,(i)}_n.\\
\end{array}
\label{eq:Vyear}
\end{eqnarray}

We now present mean-field equations for our model with the second public health program. Dividing Eqs.~\eqref{eq:Nyear} by $N$ we get a set of equations for $\eta_n^r= N_n^r/N$ which are intensive quantities. Following the technique in Sec.~\ref{sec:2}, we average over the individuals to obtain the mean-field variables. We arrive to the following dynamical system
\begin{eqnarray}
	\eta_{n+1}^0&=&\eta_n^1+(1-\pi^0_n)\eta^0_n,\\
	\label{eq:etayear0}
	\eta_{n+1}^r&=&\eta_{n}^{r+1};\,\,\mbox{for $0 <r<y$}, \\
	\label{eq:etayearr}
	\eta_{n+1}^y&=&\pi^0_n \eta^0_n,
	\label{eq:etayeary}
\end{eqnarray}
\begin{eqnarray}
	v^0_{n+1}&=&s v^0_n+(1-\pi^0_n)q(p_n)+\pi^0_n[1-\theta(p_n-p_c)],\\
	\label{eq:v0year}
	v^r_{n+1}&=&s v^r_n+1-\theta(p_n-p_c);\,\, \mbox{for $0<r\leq y$},
	\label{eq:vryear}
\end{eqnarray}
\begin{eqnarray}
\pi^0_{n+1}&=&(1-s) v^0_{n+1}/(1-s^{n+1}),\\
\label{eq:pi0year}
\pi^r_{n+1}&=&1;\,\, \mbox{for $0<r\leq y$},
\label{eq:piryear}
\end{eqnarray}
where
\begin{eqnarray}
p_n\equiv \sum_{r=0}^y \eta^r_n \pi_n^r,
\label{eq:pyear2}
\end{eqnarray}
and
\begin{eqnarray}
\sum_{r=0}^y \eta^r_n =1.
\label{eq:eta2}
\end{eqnarray}
The dynamical system simplifies since the equations for $v^r\,\,(0<r\leq y)$ and $\pi^r\,\,(0\leq r\leq y)$ are decoupled, and $\eta^0$ can be eliminated using the constraint \eqref{eq:eta2}.  We thus obtain
\begin{eqnarray}
\label{eq:etaryearr}
\eta_{n+1}^r&=&\eta_{n}^{r+1};\,\,\mbox{for $0 <r<y$},\\
\label{eq:etayyearr}
\eta_{n+1}^y&=&v^0_n\left(\frac{1-s}{1-s^n}\right)\left(1-\sum_{r=1}^y \eta^r_n \right),\\
\label{eq:v0yearr}
v^0_{n+1}&=&s v^0_n+\left[1-v^0_n\left(\frac{1-s}{1-s^n}\right)\right]q(p_n)+v^0_n\left(\frac{1-s}{1-s^n}\right)[1-\theta(p_n-p_c)],
\end{eqnarray}
where
\begin{eqnarray}
p_n\equiv F(\eta^1_n, ..., \eta^y_n,v^0_n)\equiv v^0_n\left(\frac{1-s}{1-s^n}\right)+\left[1-v^0_n\left(\frac{1-s}{1-s^n}\right)\right]\sum_{r=1}^y \eta^r_n.
\label{eq:pyear2_}
\end{eqnarray}
We write the state of the system as $(\eta^1_n, ..., \eta^y_n,v^0_n)$; the domain of the dynamical system is $\mathcal{D}=[0,1]^y\times[0,1/(1-s))$.

{\it Fixed Point Analysis.} Due to the discontinuities at $p=p_c$, we distinguish two complementary domains: $\mathcal{D}_1=\{(\eta^1, ..., \eta^y,v^0)|F(\eta^1, ..., \eta^y,v^0)\geq p_c\}$ and $\mathcal{D}_2=\{(\eta^1, ..., \eta^y,v^0)|F(\eta^1, ..., \eta^y,v^0)< p_c\}$.
 
{\bf Case 1: $\mathcal{D}_1$.} In the limit $n\rightarrow\infty$, Eqs.~\eqref{eq:etaryearr}, \eqref{eq:etayyearr} and \eqref{eq:v0yearr} become
\begin{eqnarray}
\eta_{n+1}^r&=&\eta_{n}^{r+1};\,\,\mbox{for $0 <r<y$},\\
\label{eq:etaryear_1}
\eta_{n+1}^y&=&v^0_n(1-s)\left(1-\sum_{r=1}^y \eta^r_n \right),\\
\label{eq:etayyear_1}
v^0_{n+1}&=&s v^0_n,
\label{eq:v0year_1}
\end{eqnarray}
which has no attractors in $\mathcal{D}_1$.  However, if we extend the domain to $\mathcal{D}$, then the system has a fixed point that belongs to $\mathcal{D}_2$ at $x^*=(0, ..., 0)$. This fixed point is an attractor since the jacobian of the dynamical system evaluated at $x^*$ has one eigenvalue equal to $s$ ($0\leq s<1$) and $j$ eigenvalues equal to $0$. This attractor is global since $v_0$ is decreasing, corresponds to the situation where no individual vaccinates, and fully characterizes the dynamics in $\mathcal{D}_1$: orbits in $\mathcal{D}_1$ will be attracted to $x^*$ until they land in $\mathcal{D}_2$. In $\mathcal{D}_2$, the orbit is iterated with a different smooth map.

{\bf Case 2: $\mathcal{D}_2$.} We now obtain the following dynamical system
\begin{eqnarray}
\eta_{n+1}^r&=&\eta_{n}^{r+1};\,\,\mbox{for $0 <r<y$},\\
\label{eq:etaryear_2}
\eta_{n+1}^y&=&v^0_n(1-s)\left(1-\sum_{r=1}^y \eta^r_n \right),\\
\label{eq:etayyear_2}
v^0_{n+1}&=&v^0_n+[1-v^0_n(1-s)]q(p_n),
\label{eq:v0year_2}
\end{eqnarray}
that has no fixed points in $\mathcal{D}_2$. However, if we extend the domain to $\mathcal{D}$, then the system has two fixed points. The first fixed point is $\tilde x=((y+1)^{-1}, (y+1)^{-1}, ..., (y+1)^{-1}, (1-s)^{-1})$ and corresponds to the situation where everybody vaccinates (i.e., $p=1$). It can be shown straightforwardly that $1+q(0)(1-s)(p_c^{-1}-1)$ and $1$ are eigenvalues of the jacobian of the system at $\tilde x$; thus, $\tilde x$ is unstable. The second fixed point of the extended system is $\hat x=(p_c/(y+1), p_c/(y+1), ..., p_c/(y+1), p_c/(1-s)/[1+y(1-p_c)])$ corresponding to $p=p_c$. The fixed point may be interpreted as follows. The population of $N$ individuals is divided into two groups.  The individuals in the first group never vaccinate while those in the second group always vaccinate. The number of individuals that vaccinate keep the vaccination coverage $p$ at $p_c$ every year and always have positive experience with influenza vaccination. This implies that the $\eta^r$ ($0<r\leq y$) values of $\hat x$ are all the same. $\eta^0=1-yp_c/(y+1)$ includes both individuals that never vaccinate and individuals that just finished a vaccination program and will enroll in a new one. The individuals who do not vaccinate  do not get infected because $p=p_c$ and always have positive experience with not vaccinating.  The characteristic equation at $\hat x$ is analytically intractable.  However, it can be established numerically that $\hat x$ is a potential attractor of the map defined for $\mathcal{D}_2$ when the domain is extended to $\mathcal{D}$. For the dynamical system describing the basic model with superimposed public health program 2 (Eqs.~\eqref{eq:etaryearr}, \eqref{eq:etayyearr}, and  \eqref{eq:v0yearr}), $\hat x$ may be attracting from $\mathcal{D}_2$ since $\hat x$ is on $\partial\mathcal{D}_2$ and the basin of $\hat x$ intersects $\mathcal{D}_2$; $\hat x$ is asymptotically approached, yet never reached.  

\begin{figure}
\begin{center}
\mbox{\epsfig{file=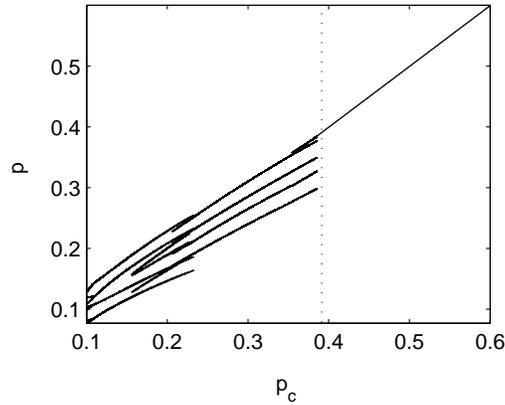,width=6.8cm}}
\end{center}
\caption{Bifurcation diagram of the dynamical system given by Eqs.~\eqref{eq:etaryearr}, \eqref{eq:etayyearr}, and  \eqref{eq:v0yearr} versus $p_c$. $s=0.7$ and the probability of getting infected when the coverage $p=0$ is $q(0)=0.8$. We slightly modified the map to reduce the effects of numerical noise; in turn, this slightly changes the bifurcation.}
\label{fig:bifdiag2}
\end{figure}

{\it Bifurcation Diagram.} Figure \ref{fig:bifdiag2} presents a bifurcation diagram for $y=1$, $s=0.7$, $q(0)=0.8$ and varying $p_c$.  At high values of $p_c$, the orbits are attracted to $\hat x$ which yields $p=p_c$. With decreasing $p_c$, a border collision bifurcation occurs. At a particular value of $p_c$, a critical periodic orbit is created in phase space which with further decreasing $p_c$ turns into a periodic attractor (see Fig.~\ref{fig:bifdiag2}). Denoting the $(y+1)$-dimensional map given by Eqs.~\eqref{eq:etaryearr}, \eqref{eq:etayyearr}, and  \eqref{eq:v0yearr} by $M$, the equation of the critical orbit is 
\begin{eqnarray}
M^{[n]}(\hat x; p_c)=\hat x,
\label{eq:critic}
\end{eqnarray}
where by $M^{[n]}$ we denoted the $n$th iterate of the map, and the solutions must have finite $n$.
Equation \eqref{eq:critic} may be understood as follows. For some parameter values, $\hat x$ is reached in a finite number of iterates. $\hat x$ ($\hat x\in\mathcal{D}_1$) is not a fixed point of the map $M$; in particular, the next iteration of $\hat x$ will be in $\mathcal{D}_2$, then the orbit will be evolved with the smooth map that applies in $\mathcal{D}_2$ for a finite number of iterates until $\hat x$ is reached again. Equation \eqref{eq:critic} may not have solutions for all positive integers; also, there may be multiple solutions with the same value of $n$. In particular, for $n=4$, $y=1$, $s=0.7$, $p_c=0.6$, and $q(0)=0.8$, we numerically solve Eq.~\eqref{eq:critic} and obtain $p_c^{th}\approx 0.4$ which is in agreement with the numerics presented in Fig.~\ref{fig:bifdiag2}.  For values of $p_c$ in the neighborhood of $p_c^{th}$ ($p<p_c^{th}$), the critical periodic orbit turns into a periodic attractor.  It is important to note that, except for $p=p_c^{th}$, the periodic attractor is bounded away from $\partial\mathcal{D}_2$, such that the periodic attractor is robust to arbitrarily small noise.  A similar analysis can be performed with varying $s$. To the best of our knowledge, this is the first time this type of border collision bifurcations is mentioned in the literature. We numerically observe that for our model, $p_c^{th}$ and $s^{th}$ increase with $y$. 

{\it Discussion.} The study of these border collision bifurcations is critical for understanding the effect of the public health program 2.  In practice, it would be unlikely that $p_c$ and/or $s$ are parameters that can be easily changed since the critical vaccination coverage $p_c$ is strongly determined by the viral strain and the memory parameter $s$ is a feature of the population.  On the other hand, $y$ would be a parameter of the public health program 2 that is easy to change.  Since $p_c^{th}$ and $s^{th}$ increase with $y$, it is expected that, for given $s$, $p_c$, and $q(0)$, there exists a threshold value of $y$ (which we denote $y^{th}$) such that the orbits of the dynamical system with parameters $y=y^{th}-1$, $s$, $p_c$, and $q(0)$ are attracted to the corresponding $\hat x$ (i.e., to $p=p_c$) and the orbits of the dynamical system with parameters $y=y^{th}$, $s$, $p_c$, and $q(0)$ go to a non-trivial period $n$ attractor.  For example, for $s=0.7$, $p_c=0.6$, and $q(0)=0.8$, we obtain $y^{th}=3$ and $n=6$.  This phenomenon is very important since it establishes a {\it threshold} value for the number of prepaid vaccinations such that the public health program 2 makes a significant difference in the dynamics of the mean field coverage.  Furthermore, depending of the parameters of the system, the statistics over time of the orbit on the periodic attractor may yield better epidemiological quantifiers that the statistics over time of the orbit attracted to $p_c$.  In this case, the public health program 2 is effective in ameliorating the influenza epidemiology.

\section{Conclusions}

We introduced (for the first time) an inductive reasoning game to model whether or not the critical vaccination coverage can be reached and influenza epidemics can be prevented by voluntary vaccination in a large population of individuals acting in their own self-interest.  From our analysis, it obtained that epidemics are only occasionally prevented.  We explored two public health programs based on offering incentives in a voluntary vaccination market.  The first public health program proposed dividing the population into families and leaving the vaccination decisions to the heads of the family.  Surprisingly, this program always exacerbated epidemics since it reduced the number of independent decision makers.  The second public health program that we analysed required prepayment of vaccinations.  We found that there exist a critical number of prepayments so that the public health program will make any difference at all for the course of the epidemics.  Once this critical number of prepayments is established and implemented, depending on the population parameters and the epidemiological parameters, the second public health program has the potential to ameliorate epidemics.

\vspace{5mm}
{\bf Acknowledgments}\vspace{5mm}

The authors gratefully acknowledge financial support from NIH/NIAID (RO1 AI041935). We thank Tiffany Head for assistance with the preparation of the manuscript.

\pagebreak

\pagebreak

\pagebreak
\newpage
\clearpage
\appendix
\section{}
\label{Appendix:A}

\begin{figure}\begin{center}
\mbox{\epsfig{file=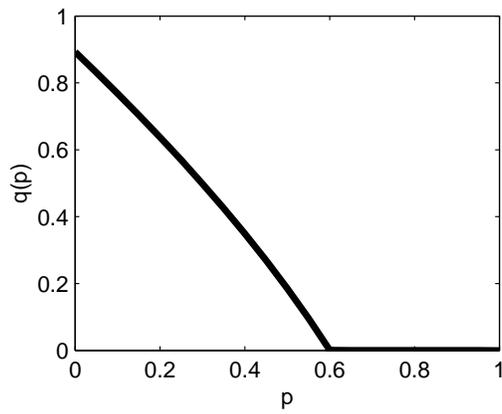,width=68mm}}\end{center}
\caption{The probability of getting infected $q(p)$ versus the vaccination coverage $p$ for the SIR model with no vital dynamics. The parameters are $N=10^5$, $\beta=5/6$ day$^{-1}$, $\gamma=1/3$ day$^{-1}$ and $T=200$ days.}
\label{fig:3.1}
\end{figure}

In order to calculate the probability of getting infected  with influenza $q$ given a certain vaccination coverage $p$ during one vaccination season we make the following assumptions:
\begin{enumerate}
	\item We ignore the inflow and outflow of individuals in the study population during a season. That is, we ignore vital dynamics. 
	\item Individuals may vaccinate against influenza only before the beginning of the influenza season.  
	\item Vaccinated individuals are not susceptible during the next coming season.
	\item Individuals who get infected and then recover remain immune to infection until the end of the season.
	\end{enumerate}
As a result of the above assumptions we choose to model the epidemic transmission during one season using a SIR model without vital dynamics that includes vaccination at the beginning of each influenza season.
\begin{eqnarray}
	\begin{array}{l}
	dS(t)/dt= -\beta S(t)I(t)/N,\\
	dI(t)/dt= \beta S(t)I(t)/N-\gamma I(t),\\
	dR(t)/dt= \gamma I(t),\\
	dV(t)/dt= 0,
	\end{array}
\end{eqnarray}
where $S(t)$, $I(t)$, $R(t)$ and $V(t)$ represent the number of susceptible, infected, recovered and vaccinated individuals, respectively. The total number of individuals $N=S(t)+I(t)+R(t)+V(t)$ is constant. $\beta$ represents the transmissibility in the mass-action term \cite{McCallum}. $\gamma$ represents the recovery rate. The probability of getting infected during an influenza season $q(p)$ is given by \begin{eqnarray}q(p)= \int_0^T \beta S(t)I(t)/N^2 dt,\end{eqnarray}
where $T$ represents the duration of the influenza season. The initial conditions are as follows. A fraction $p$ of the population vaccinates against influenza leaving only $(1-p)N$ susceptibles. Thus, at the start of the influenza season, $S(0)=(1-p)N-1$, $I(0)=1$, $R(0)=0$ and $V(0)=pN$.  In Fig.~\ref{fig:3.1} we show a illustrative graph of $q(p)$. We note that the featured dependence is approximately piecewise linear. The discontinuity in derivative occurs at $p=p_c=1-1/N-\gamma/\beta$ which for large $N$ becomes $p_c\approx 1-\gamma/\beta$ \cite{AM}.

\end{document}